# Tuning epitaxial graphene sensitivity to water by hydrogen intercalation

C. Melios[1,2], M. Winters[3], W. Strupiński[4], V. Panchal[1], C. E. Giusca[1], K.D.G.I. Jayawardena[2], N. Rorsman[3], S.R.P. Silva[2], and O. Kazakova[1*]

[1]*National Physical Laboratory, Teddington, TW11 0LW, UK*

[2]*Advanced Technology Institute, University of Surrey, Guildford, GU2 7XH, UK*

[3]*Chalmers University of Technology, Dept. of Microtechnology and Nanoscience, Göteborg, 412-96, Sweden*

[4]*Institute of Electronic Materials Technology, Warsaw, 01-919, Poland*

**Abstract**

The effects of humidity on the electronic properties of quasi-free standing one layer graphene (QFS 1LG) are investigated via simultaneous magneto-transport in the van der Pauw geometry and local work function measurements in a controlled environment. QFS 1LG on 4*H*-SiC(0001) is obtained by hydrogen intercalation of the interfacial layer. In this system, the carrier concentration experiences a two-fold increase in sensitivity to changes in relative humidity as compared to the as-grown epitaxial graphene. This enhanced sensitivity to water is attributed to the lowering of the hydrophobicity of QFS 1LG, which results from spontaneous polarization of 4*H*-SiC(0001) strongly influencing the graphene. Moreover, the superior carrier mobility of the QFS 1LG system is retained even at the highest humidity. The work function maps constructed from Kelvin probe force microscopy also revealed higher sensitivity to water for 1LG compared to 2LG in both QFS 1LG and as-grown systems. These results point to a new field of applications for QFS 1LG, *i.e.,* as humidity sensors, and the corresponding need for metrology in calibration of graphene-based sensors and devices.

## 1. Introduction

In recent years, graphene has received significant attention due to its exceptional electronic properties[1]. In particular, owing to its extraordinary surface-to-volume ratio together with mechanical robustness, graphene holds a great promise for application in the sensor industry[2]. Growth of epitaxial graphene on SiC has proven to be a promising method for the production of wafer-scale graphene[3–5], which can be used in novel high-speed analogue transistors, due to its high charge carrier mobility[6]. Despite that, the electronic properties (work function, carrier concentration and mobility) of graphene on SiC depend highly on the substrate, surrounding environment and number of layers[7,8]. For example, graphene on 4*H*-

SiC(0001), *i.e.* Si face of SiC, exhibits electron conduction due to the charge transfer from the interfacial layer (IFL)[9]. IFL is a $(6\sqrt{3} \times 6\sqrt{3})R30°$ reconstructed carbon layer, topographically similar to graphene. It consists of a mixture of $sp^2$ and $sp^3$ carbon atoms, with a significant fraction of them bonded to the SiC substrate[3]. This carbon layer is still covalently bonded to the substrate, thus acting as a source of impurity and phonon scattering centre, which is the main reason for the charge carrier mobility degradation, as observed in this type of graphene[9]. Moreover, the level of intrinsic doping in epitaxial graphene layers (on SiC(0001)) is thickness dependent[4,10] and can be further extrinsically doped by airborne adsorbates[8] (such as $O_2$ and $NO_2$) and ambient humidity[7,8]. This extrinsic doping due to atmospheric dopants is found to be also strongly thickness dependent [7,8].

A favored route to achieve decoupling of the graphene from the SiC substrate is by hydrogen intercalation. Several groups have successfully demonstrated decoupling of the IFL and its transformation into quasi-free standing graphene (QFSG) [9,11–13]. While the principle of intercalation is very simple, in practice achieving the optimum result is challenging, as graphene may be either only partially decoupled or alternatively etched. In theory, by annealing the graphene sample at high temperatures (550 – 1100 °C, with the precise temperature being highly dependent on the growth system) in a hydrogen environment, the hydrogen will penetrate underneath the graphene and break the Si-C bonds[9,11–13]. Following this, hydrogen will passivate the Si substrate and create Si-H bonds. The formation of Si-H bonds, which demonstrates the successful passivation of the SiC substrate with hydrogen, was demonstrated previously using surface enhanced Raman scattering (SERS)[14], Fourier transform infrared spectroscopy (FTIR)[9] and X-ray photoelectron spectroscopy (XPS)[11]. This forces the IFL to decouple from the SiC substrate and be converted to a QFSG. The main result of this transformation is the change of the intrinsic doping from electron to hole type (due to the spontaneous polarization (SP) of the SiC substrate) and the significant enhancement of carrier mobilities (due to effective decoupling of graphene from the substrate)[10,15,16]. While QFSG had already resulted in improved intrinsic cut-off frequency graphene field-effect transistors (GFET)[6] and low-noise Hall effect sensors[17], there are currently no studies investigating the changes in its electronic properties due to atmospheric influence, in particular changes in humidity. Since water is the most abundant dipole adsorbate in the ambient air, it is important to investigate its effects on QFSG. Understanding the influence of water vapour on QFSG will lead to the development of stable graphene-based electronics, such as GFETs and appropriately calibrated sensors, by minimizing or accounting for the effects of humidity. Furthermore, as carrier concentration and mobility are used as a figure of merit for graphene quality, it is important for standardization procedures to unambiguously specify the measurement conditions and carrier concentration for each reported mobility value, as environmental conditions such as humidity and temperature can influence the measurement[18]. Such carrier concentration-mobility benchmark curve was initially presented by Dimitrakopoulos *et al.*[19,20]. This is particularly important for future graphene-based electronics, which will operate under different environmental conditions (*i.e.*, from vacuum to variable humidity). As an example, the minimum detectable magnetic field as measured by Hall effect in QFSG strongly depends on the carrier concentration[21–23], thus any variations in humidity can affect the sensor performance (if devices are not encapsulated).

In this work, we employ simultaneous Kelvin probe force microscopy (KPFM) and transport measurements in the van der Pauw geometry in order to correlate the changes in the local and global electronic properties of graphene. The study has been performed both on as-grown graphene on SiC and quasi-free standing one layer graphene (QFS 1LG), in an attempt to compare the effects of the water vapour and their interplay with substrate-induced effects, rather than a general study of environment induced doping. KPFM is a powerful technique, which maps the surface potential of the graphene sample with nanometre resolution, providing local information about the work function and layer thickness, while the transport measurements offer complimentary global information about the charge carrier concentration and mobility. By performing the measurements in a well-control environment with relative humidity (R.H.) ranging from 0 to 80%, we are able to monitor the simultaneous changes in surface potential, work function, charge carrier concentration and mobility. These measurements provide the essential information on the mechanisms of graphene sensitivity to water, specifically underlying the role of water molecule – graphene – substrate interactions and enabling the appropriate calibration of gas and humidity sensors, highlighting the need for encapsulation of other graphene-based devices and stress the importance of standardisation of carrier concentration and mobility measurements under specific environmental conditions.

## 2. Experimental section

### 2.1. Sample growth

Two types of samples were investigated: 1) as-grown and 2) QFS 1LG. The samples were grown by CVD method at 1600 °C under a laminar flow of argon with a few ppm methane in an Aixtron VP508 hot-wall reactor. Semi-insulating on-axis oriented 4*H*-SiC(0001) substrates (Cree) of 10×10 mm$^2$ size were cut out from 4" wafer and etched in hydrogen at 1600 °C prior to the epitaxy process. Graphene growth was controlled by Ar pressure, Ar linear flow velocity and reactor temperature. The process relies critically on the creation of dynamic flow conditions in the reactor, which control Si sublimation rate and enable the mass transport of methane (precursor) to the SiC substrate. Tuning the value of the Reynolds number enables the formation of an Ar boundary layer thick enough to prevent Si sublimation and allowing the diffusion of hydrocarbon to the SiC surface, followed by epitaxial CVD growth of graphene on the SiC surface[24]. For the growth of the IFL, annealing of the substrate was terminated just before 1LG was formed[12]. *In-situ* intercalation of hydrogen was achieved by annealing the IFL sample in hydrogen at temperature of ~1100 °C and reactor pressure of 900 mbar. The success of the intercalation process was previously evaluated using surface enhanced Raman spectroscopy/mapping (where the IFL was transformed into the 1LG)[16], which *e.g.* demonstrated the appearance of the Si-H peak at ~2100 cm$^{-1}$,[14] and magneto-transport measurements showing the change of the sample conduction from n- to p-type[12,14,16].

### 2.2. Van der Pauw device fabrication

In order to investigate the low field transport properties of the graphene layers, 5 μm × 5 μm symmetric van der Pauw structures were fabricated using an electron beam lithography (EBL) process. All lithography steps were performed in a JOEL 9300FS electron beam lithography system, and all metallization steps were performed by electron beam evaporation. Following the deposition of alignment marks, fabrication begins with a mesa isolation step. Lithography was performed in a negative mode, and the graphene was etched using a 40 W $O_2$ plasma for 30 s. Ohmic contacts were then patterned using positive resist and subsequent lift-off of a Ti(1 nm)/Pd(30 nm) metallization stack. After metallization, the contacts were annealed at 400 °C under an argon atmosphere for 10 min. This annealing step serves to promote adhesion of the Ohmic contact layer to the graphene. Finally large contact pads were defined in a positive mode and lift-off process using Ti(10 nm)/Au(100 nm) as metallization.

*2.3. Transport measurements in the van der Pauw geometry*

An AC transport measurement system in the van der Pauw geometry compatible with SPM setup and environmental chamber was developed and allowed for carrier concentration measurements on the 5×5 μm² graphene device (figure 1). The sample was placed in an electromagnet creating a magnetic field of $B_{AC}$=5 mT (peak to peak). The sheet resistance ($R_S$) was measured using a digital voltmetre (DVM) and calculated using $e^{\frac{-\pi R_A}{R_S}} + e^{\frac{-\pi R_B}{R_S}} = 1$, where $R_A$ and $R_B$ are the resistances obtained by passing a bias current $I_B$ = 100 μA and measuring the voltage drop across the opposite sides of the sample. To obtain the carrier concentration ($n = \frac{1}{eR_H}$), the Hall coefficient ($R_H$) was measured using $R_H = \frac{V_H}{BI}$, by passing current and measuring the diagonal AC Hall voltage ($V_H$) of the sample using a lock-in amplifier (LIA). The mobility was also calculated using $\mu = \frac{R_H}{BR_S}$ [25].

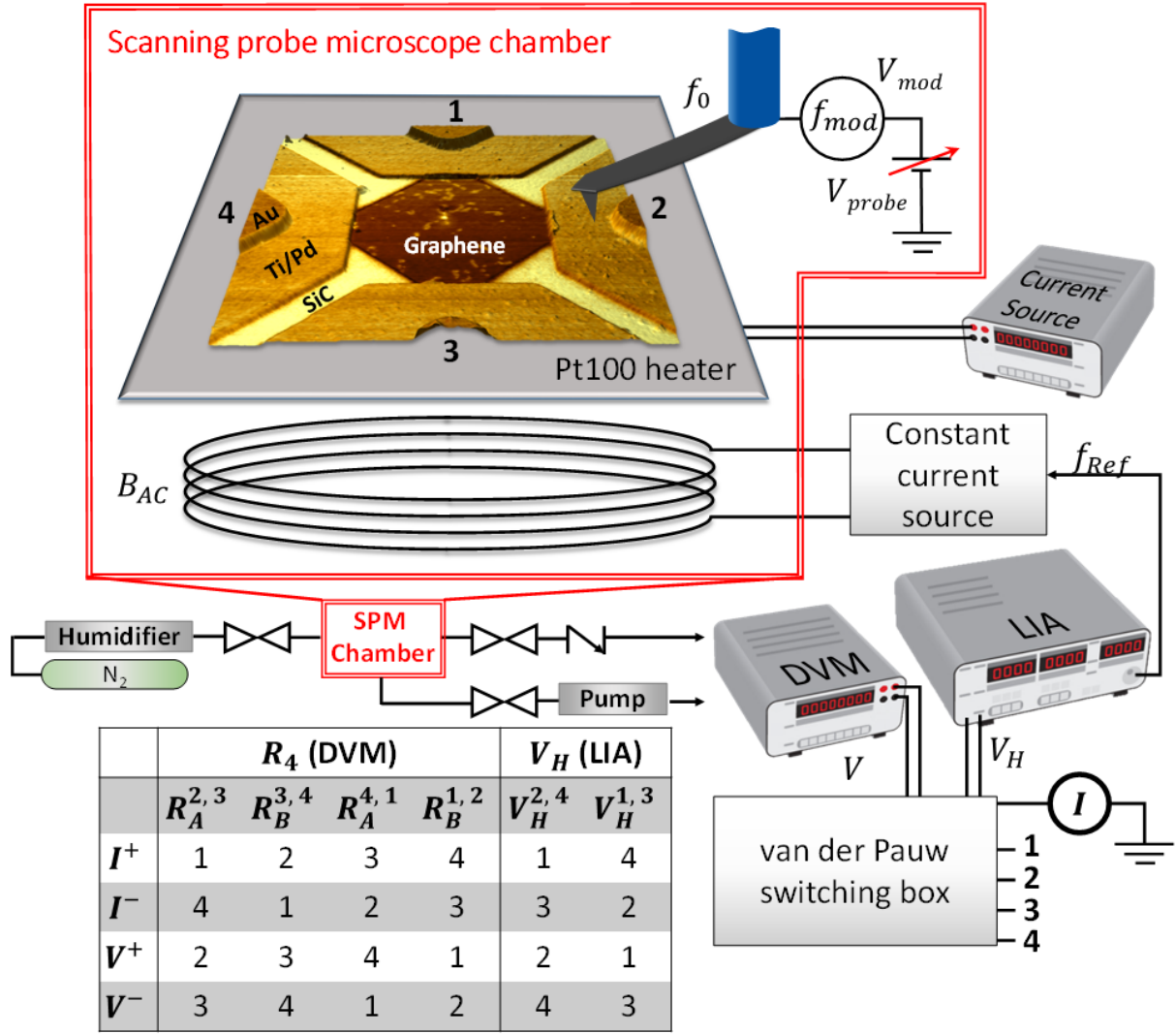

**Figure 1:** Schematic of the experimental set-up used to measure surface potential (work function) and transport characteristics in the environmental SPM chamber. The red contour shows the environmental enclosure. The transport properties in the van der Pauw geometry were measured using a custom-made switching box, allowing for the different contact combinations as seen in the included table. During surface potential scanning the transport measurements were paused and the sample was grounded.

## 2.4. Frequency-modulated Kelvin probe force microscopy

A NT-MDT Ntegra Aura SPM system equipped with an environmental chamber was employed in single pass frequency-modulated KPFM (FM-KPFM) mode (figure 1). In this tapping mode, a doped silicon probe (PFQNE-AL) with spring constant $k \approx 0.4$–$1.2$ N m$^{-1}$ is oscillating at its mechanical resonant frequency $f_0 \approx 300$ kHz, while a much lower frequency ($f_{mod} \approx 3$ kHz) modulating AC voltage is applied to induce a frequency shift of $f_0 \pm f_{mod}$. The side lobes (monitored by a PID feedback loop) generated by this shift are minimized by applying a DC compensation voltage. By measuring the DC voltage at each pixel, a surface potential map [contact potential difference ($U_{CPD}$)] is constructed. As FM-KPFM is a force gradient technique, a high spatial resolution of <20 nm can be achieved, which is limited only by the tip apex diameter[26,27]. This allows nanometre resolution imaging of the surface potential of graphene and provides direct information of the work function variations and number of layers[16,28–30]. For the calculation of the tip work function at each environmental stage, we used the gold contacts ($\Phi_{Au} = 4.9\ eV$[16]) of the van der Pauw device, as a reference, where the work function of the probe was obtained using the equation: $\Phi_{Tip} = \Phi_{Au} + eU_{CPD}$. Then, the work function of the graphene was calculated using $\Phi_{Sample} = \Phi_{Tip} - eU_{CPD}$.

*2.5. Environmental control*

For the investigation of the effects of humidity on the electronic properties of as grown and QFS 1LG, experiments were performed in a controlled environmental scanning probe microscope (SPM) chamber, by monitoring the global carrier concentration and local surface potential, with atmospheric conditions changing in the following order: ambient conditions (~50 °C, R.H. ~35%), vacuum (P=1×10$^{-5}$ mbar), dry nitrogen (research grade 99.9995% purity), gradually increasing humidity level (R.H. = 0-80%, with measurements taken at steps of 20% R.H. ) balanced with nitrogen and finally the sample was brought back to the ambient conditions. It is noteworthy that heating of the electromagnet coil caused higher (but constant) temperature than room temperature (*i.e.* ~50ºC). Prior to the experiment, residual polymers were removed from the graphene area using contact mode atomic force microscope (AFM).

*2.6. Water contact angle measurements*

Water contact angle (WCA) measurements were performed in ambient conditions by depositing 120 μL of water on the non-patterned graphene samples using a Kruss EasyDrop system. Both the as-grown and QFS 1LG samples were cleaned by vacuum annealing (~12 hours) and kept in vacuum prior the measurements to minimize any surface contamination.

**3. Results and discussion**

*3.1. Effects of humidity on transport properties of as-grown graphene*

In ambient, the as-grown sample exhibits electron conduction ($n_e$=3.1×10$^{12}$ cm$^{-2}$), due to the IFL being a constant source of electrons. Furthermore, the IFL acts as a strong source of scattering, limiting the mobility to $\mu_e$=866 cm$^2$/Vs for this particular sample. Following annealing (~160 ºC) in vacuum (P=1×10$^{-5}$ mbar) for ~10 hours, the sample reaches its intrinsic state (*i.e.* all airborne dopants are desorbed) with maximum increase of the electron concentration ($n_e$=1.18×10$^{13}$ cm$^{-2}$) but only marginal change of mobility, $\mu_e$=870 cm$^2$/Vs. At this stage, two competing mechanisms affect mobility. Firstly, increase in electron concentration results in increased electron–electron interaction, thus lowering the mobility. Additionally, desorption of water and other adsorbates from the surface results in lower impurity scattering, therefore the mobility tends to increase. As the result of this competition, mobility remains relatively constant at this stage. This shows that vacuum annealing can be used effectively for desorption of airborne contaminants from the graphene surface, effectively increasing the electron concentration. The summarized measurements are presented in Figure 2a, bottom panel.

Following vacuum annealing to ensure the clean state of the sample, dry nitrogen was introduced. Despite the inert nature and the high purity of $N_2$, there is still a notable decrease in the electron concentration, which is possibly caused by impurities transferred from the plastic pipes (regardless of intense flushing). Table 1 shows the absolute changes of carrier concentration and mobility with respect to the vacuum level (pristine state of sample) and after exposure to nitrogen (which is used as a reference value). Following dry nitrogen, 20% R.H. was added in the chamber, which decreased the electron concentration by $0.35\times10^{12}$ cm$^{-2}$, while the mobility was increased by 4%, compared to the nitrogen state (Table 1). This indicates a mild p-type doping of graphene by water vapor. Interestingly, despite the further increase in humidity (up to 80% R.H.), the electron concentration shows only a moderate decrease, reaching $\sim n_e=1.01\times10^{13}$ cm$^{-2}$ or 7.2% reduction from the initial level in nitrogen. The change in electron concentration and mobility as a function of humidity is presented in figure 2a, bottom panel. Figure 2d also shows the absolute values of carrier mobility as a function of carrier concentration, plotted for various humidity levels. Regarding the carrier mobility, an initial increase was observed at 20% R.H. (figure 2a) due to decreased electron concentration and partial neutralization of the Coulomb scattering centers in the IFL by water reducing their scattering potentials[8]. However, further increase in humidity (up to 80% R.H.) resulted in a linear decrease in mobility. Considering the overall trend of mobility (figure 2d) to decrease with humidity (accompanied by decrease of the electron concentration), we suggest that the mobility changes mainly due to Coulomb and impurity scattering owed to the presence of a water layer on the graphene surface.

**Table 1:** Summary of the humidity induced changes of carrier concentration ($\times10^{12}$ cm$^{-2}$) and mobility (cm$^2$/Vs) in respect to the vacuum and nitrogen states. ↑/↓ symbols notified the increased/decreased values as compared to the vacuum or nitrogen stage.

|  | Difference with vacuum stage | | | | Difference with nitrogen stage | | | |
|---|---|---|---|---|---|---|---|---|
| *Stage* | $\Delta n_e^{As\text{-}grown}$ | $\Delta \mu_e^{As\text{-}grown}$ | $\Delta n_h^{QFS}$ 1LG | $\Delta \mu_h^{QFS}$ 1LG | $\Delta n_e^{As\text{-}grown}$ | $\Delta \mu_e^{As\text{-}grown}$ | $\Delta n_h^{QFS}$ 1LG | $\Delta \mu_h^{QFS}$ 1LG |
| **20% R.H.** | ↓ 1.15 | ↑ 103 | ↑ 1.7 | ↓ 1117 | ↓ 0.35 | ↑ 39 | ↑ 0.76 | ↓ 102 |
| **40% R.H.** | ↓ 1.26 | ↑ 101 | ↑ 2.05 | ↓ 1154 | ↓ 0.46 | ↑ 37 | ↑ 1.11 | ↓ 139 |
| **60% R.H.** | ↓ 1.41 | ↑ 91 | ↑ 2.31 | ↓ 1178 | ↓ 0.61 | ↑ 27 | ↑ 1.37 | ↓ 163 |
| **80% R.H.** | ↓ 1.63 | ↑ 78 | ↑ 2.54 | ↓ 1193 | ↓ 0.83 | ↑ 14 | ↑ 1.60 | ↓ 178 |

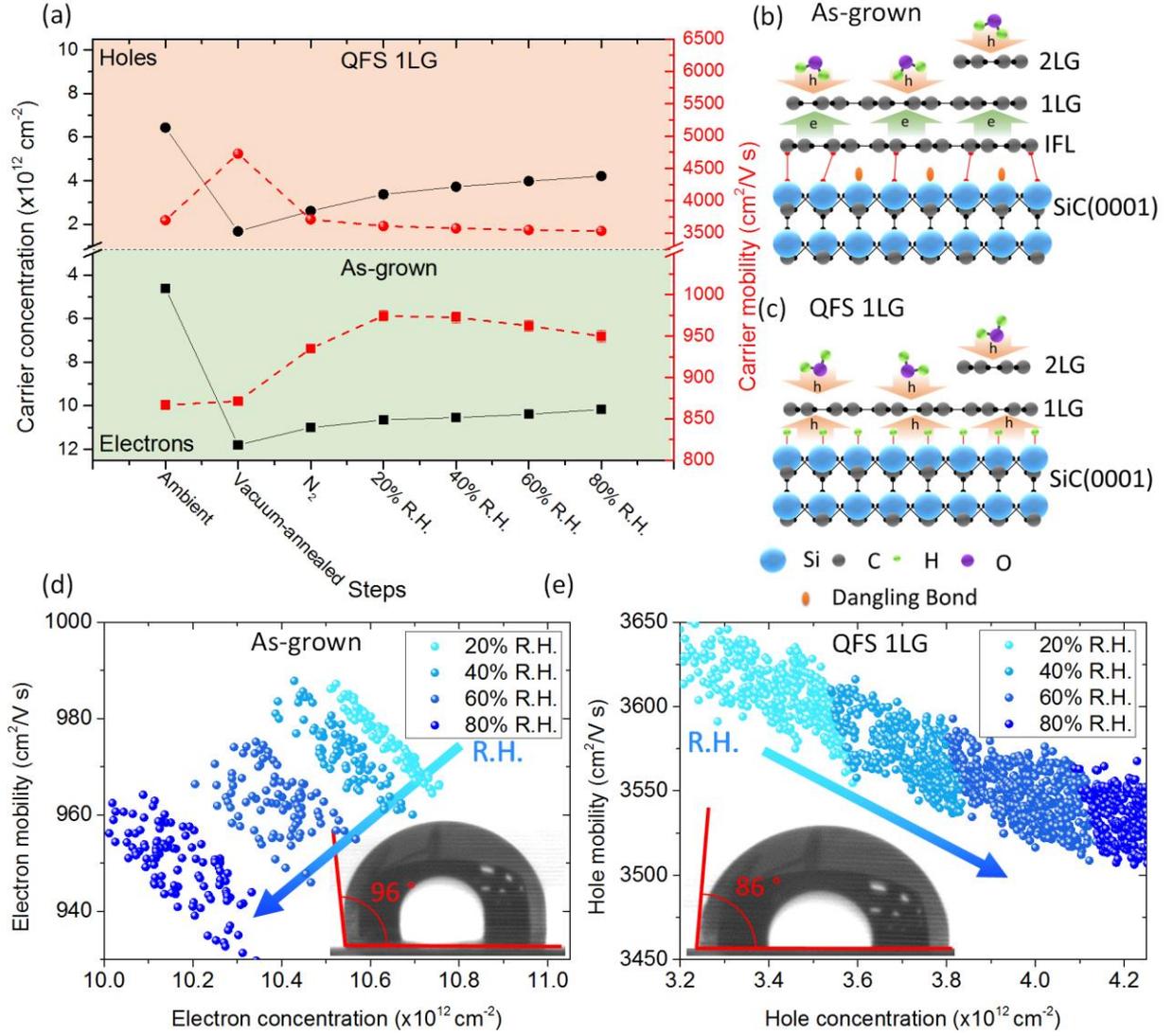

**Figure 2:** (a) Carrier concentration (black) and mobility (red) at various humidity levels for as-grown (squares, bottom panel) and QFS 1LG (circles, top panel). (b-c) Schematic representation of the graphene structure and charge transfer for (b) as-grown and (c) QFS 1LG. (d-e) Carrier mobility as a function of carrier concentration for (d) as-grown and (e) QFS for various humidity levels. Insets in (d-e) show the water contact angle measurements performed in ambient conditions.

### 3.2. Effects of humidity on transport properties of QFS 1LG

Magneto-transport measurements were also performed on QFS 1LG on 4$H$-SiC(0001) in order to be compared to the as-grown sample. A schematics of the transformation of as-grown graphene to QFS 1LG is shown in figure 1 b and c, where decoupling of the IFL and its successful transformation to QFS 1LG result in a change of carrier type (*i.e.* electron to hole), accompanied by a significant increase in mobility as compared to the as-grown sample, *i.e.* $n_h$=6.43×10$^{12}$ cm$^{-2}$ and $\mu_h$=3700 cm$^2$/Vs in ambient (figure 1a)[20]. Such high mobilities are extremely advantageous for the development of high-speed analogue transistors. Similarly to the as-grown sample, in order to achieve an intrinsic state, the sample was annealed (~160 ºC) in vacuum (P=1×10$^{-5}$ mbar) for ~14 hours, which resulted in the decrease in hole concentration and increase of mobility to $n_h$=1.67×10$^{12}$ cm$^{-2}$ and $\mu_h$=4726 cm$^2$/Vs, respectively.

The effects of humidity on QFS 1LG were studied using transport measurements. Subsequently to cooling down to 50 ºC in vacuum, dry nitrogen was introduced into the chamber (where increase in hole concentration was observed, similarly to the as-grown sample), followed by 20% R.H., which resulted in a pronounce increase of the hole concentration and decrease of the hole mobility to $n_h$=3.37×10$^{12}$ cm$^{-2}$ and $\mu_h$=3600 cm$^2$/Vs, respectively (figure 2a). Interestingly, QFS 1LG shows greater change in carrier concentration compared to the as-grown sample, when humidity is introduced (0.76×10$^{12}$ cm$^{-2}$ increase for the QFS 1LG, compared to 0.35×10$^{12}$ cm$^{-2}$ decrease for the as-grown at 20% R.H.). The humidity in the chamber was further increased until it reached a maximum of 80% R.H., resulting in the progressively increased hole concentration. In contrast with the as-grown sample, where the total change of humidity (80% R.H.) resulted in 0.83×10$^{12}$ cm$^{-2}$ decreased in electron concentration, the QFS 1LG sample exhibits a total of 1.6×10$^{12}$ cm$^{-2}$ increase in hole concentration (Table 1).

Relating the change (compared to vacuum or nitrogen environments) in carrier concentration between the two graphene types, the QFS 1LG exhibits a two-fold larger increase in sensitivity to water compared to the as-grown graphene, when the sample is exposed at 20-80% R.H. The increase sensitivity indicates that both the substrate and substrate-induced doping play a crucial role in the effective doping of graphene by water. Wehling et al.[31] demonstrated that in the case of graphene on SiO$_2$, substrate defects enable influence of water on the electronic properties of graphene, while Ashraf et al.[32] tuned the wettability of graphene on SiO$_2$ by engineering the interface between substrate and graphene. Furthermore, Hong et al.[33] demonstrated that the graphene affinity to water can be strongly affected by the substrate-induced doping, with p-doped graphene being more hydrophilic than n-type, due to the change in the water molecule orientation with the graphene doping. Our results suggest the following possible mechanism of the enhanced sensitivity of QFS 1LG compared to as-grown. SP occurs in dielectric crystals where the stacking sequence is altered (such as hexagonal polytypes of SiC, i.e. 4H or 6H). When the surface translational symmetry of the alternative stack layers breaks, the individual dipoles of each stacked layer add up resulting in a polarization field, generating a surface negative pseudo-charge. This negative pseudo-charge results in depletion of the electrons in the QFS 1LG and therefore p-doping of the graphene[10,15,34] (figure 2c). This p-doping of the QFS 1LG allows the water molecule to orient at a different angle (compared to the electron doped as-grown graphene). As a result, the graphene exhibits lower hydrophobicity and therefore higher effective doping (as also demonstrated by Hong et. al. for n- and p-type graphene on SiO$_2$[33]). In order to support the hypothesis of QFS 1LG being less hydrophobic compared to as-grown graphene, we performed WCA measurements in ambient conditions. Representative images of the WCA measurements (insets in figure 2d and e) confirm that p-type QFS 1LG is less hydrophobic than n-type as grown graphene, demonstrating 86° and 96° average contact angle values, respectively. Although it is evident that the change of doping type due to the modification of the graphene-SiC interface is responsible for the enhanced sensitivity of QFS 1LG to water, we cannot rule out additional mechanisms related to the hydrogen layer underneath the graphene. Lastly, considering the large difference in hole concentration between 80% R.H. and ambient, it is suggested that additional airborne contaminants affect the doping in ambient air (see also Supplementary Information).

Additionally, QFS 1LG exhibits notable changes in the carrier mobility in the presence of water molecules. Figure 2e plots the raw data values for carrier mobility as a function of carrier concentration, as measured for various humidity levels. Here, the hole mobility exhibits a decrease with increasing hole concentration and humidity levels (reaching $\mu_h$=3533 cm$^2$/Vs, when ~80% R.H. is introduced into the chamber). This effect may be attributed to two potential scattering mechanisms. Firstly, increase in hole concentration due to the p-doping of water results in charge carrier scattering. Furthermore, similarly to the as-grown sample, water introduces a short-range impurity scattering, thus lowering the mobility with increased humidity levels. In order to clarify the efficiency of each individual scattering mechanism, additional experiments involving gating of the devices are required.

### 3.3. Effects of humidity on local electronic properties of as-grown graphene

We further investigate how humidity affects local electronic properties of graphene. A work function map of the n-type as-grown graphene in ambient conditions is presented in Figure 3a. The sample features predominantly 1LG (light contrast, ~73%), with 2LG island inclusions (dark contrast). The assignment of each contrast level with a certain layer number was achieved using Raman spectroscopy and mapping as explained in previous works[14,16]. Both the substrate quality (*e.g.* point defects and dislocations on the SiC substrate) and the growth conditions result in faster kinetics for graphene nucleation and formation of 2LG islands[35].

In the as-grown graphene, the work function difference ($\Delta\Phi$) between 2LG and 1LG in ambient is $\Delta\Phi_{2-1LG} = -140\ meV$ (Figure 3b). This translates that 1LG has higher work function ($\Phi$) compared to 2LG, as $U_{CPD}$ is directly related to work function[a].

Following measurements in ambient, the sample was annealed in vacuum at ~160 °C for ~10 hours, which promoted desorption of physisorbed atmospheric adsorbates, therefore allowing for measurements of the intrinsic graphene properties. The control state of the sample was evaluated by a number of subsequent experiments, where the carrier concentration repeatedly reached the same level ($n_e$≈1.18×10$^{13}$ cm$^{-2}$) in vacuum. For simplicity we only display the work function map obtained in ambient conditions, as identical experiments for as-grown graphene are reported in the previous work by Giusca et al.[7]. Upon cooling to room temperature, the work function difference, $\Delta\Phi$, changes the sign, i.e. $\Delta\Phi_{2-1LG} = 110\ meV$, indicating that the 1LG has lower work function compared to 2LG. The reversal of work function between 1LG and 2LG was previously explained accounting for the difference in hydrophobicity between the two layers, where 1LG is more hydrophilic compared to 2LG[7,36]. The significant increase of the work function difference in vacuum is consistent with desorption of p-dopants, which are loosely attached to the graphene surface as also observed in previous experiments[7,8,37].

---

[a] This does not necessarily imply higher electron concentration, as the $E_F$-$n$ relation of *AB*-stacked graphene at low energies, has a more complicated relation compared to $E_F^{1LG} = v_F \hbar \sqrt{\pi n_e}$. For low carrier concentrations in *AB*-stacked graphene ($< 5 \times 10^{12}$ cm$^{-2}$), the band structure can be considered as parabolic, with $E_F^{2LG} = \pi \hbar n_e / 2 m_e^*$ (where $E_F$ is the Fermi energy, $v_F$ the Fermi velocity and $m_e^*$ the electron's effective mass, which depends on carrier concentration), while for larger concentrations the dispersion becomes linear[39]. However, the exact $n_{1LG}$ and $n_{2LG}$ values cannot be explicitly measured using the given sample geometry.

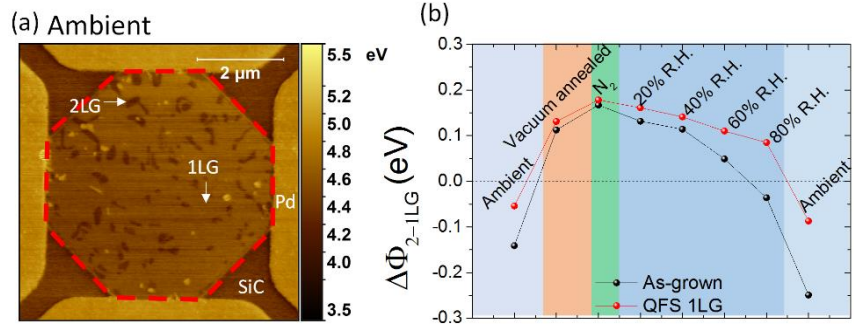

**Figure 3:** (a) Work function map of the as-grown sample in ambient. The active area of the graphene van der Pauw structure (enclosed in red dashed lines) features 1LG (light) background with 2LG island inclusions (dark). (b) Summary of the work function difference between 1LG and 2LG for the as-grown (black) and QFS 1LG (red) samples in different environmental conditions.

Subsequent to the measurements in vacuum, dry nitrogen was introduced into the SPM chamber in order to bring the chamber to atmospheric pressure. Despite the inert nature and the high purity of $N_2$, there is a notable increase in the work function difference of $\Delta\Phi_{2-1LG} = 170\ meV$. Subsequently, the chamber was brought to atmospheric pressure and water vapour (balanced with dry nitrogen) was introduced into the chamber in a controlled manner. At initial introduction of 20% R.H., the work function difference between 1LG and 2LG to $\Delta\Phi_{2-1LG} = 130\ meV$. Further increase of humidity (up to 80% R.H.), results in reversal in the work function difference to $\Delta\Phi_{2-1LG} = -40\ meV$, but regardless of much higher humidity levels (compared to ambient) $\Delta\Phi_{2-1LG}$ does not reach the ambient levels. This demonstrates that humidity is partially responsible for the higher work function of as-grown 1LG compared to 2LG observed in ambient air, but other p-dopants are also responsible for the high p-doping of the sample measured in ambient conditions[8]. Although these effects are not a part of this study, additional measurements of the doping caused by $O_2$, $CO_2$ and $NO_2$ are presented in Supplementary Information. However, upon re-exposure to ambient, 1LG shows higher work function compared to 2LG, with $\Delta\Phi_{2-1LG} = -250\ meV$.

*3.4. Effects of humidity on local electronic properties of QFS 1LG*

A sequence of the work function measurements for QFS 1LG and their summary are presented in figure 4. Figure 4a shows the work function map of the QFS 1LG sample in ambient conditions, which is covered by predominantly 85% 1LG (lighter contrast) with 2LG island inclusions (darker contrast). Similarly to the case of as-grown graphene, during the initial growth of IFL certain areas (terrace edges and pits) promote faster growth, resulting in formation of 1LG and subsequently in thicker graphene following intercalation. In ambient conditions, the work function difference ($\Delta\Phi$) between 1LG and 2 LG is $\Delta\Phi_{2-1LG} = -50\ meV$. Upon annealing (~160 ºC) and cooling down the sample, the work functions for 1LG and 2LG are $\Phi_{1LG} = 4.79\ eV$ and $\Phi_{2LG} = 4.92\ eV$, respectively, resulting in a contrast reversal between 1LG and 2LG (figure 4b) with $\Delta\Phi_{2-1LG} = 130\ meV$.

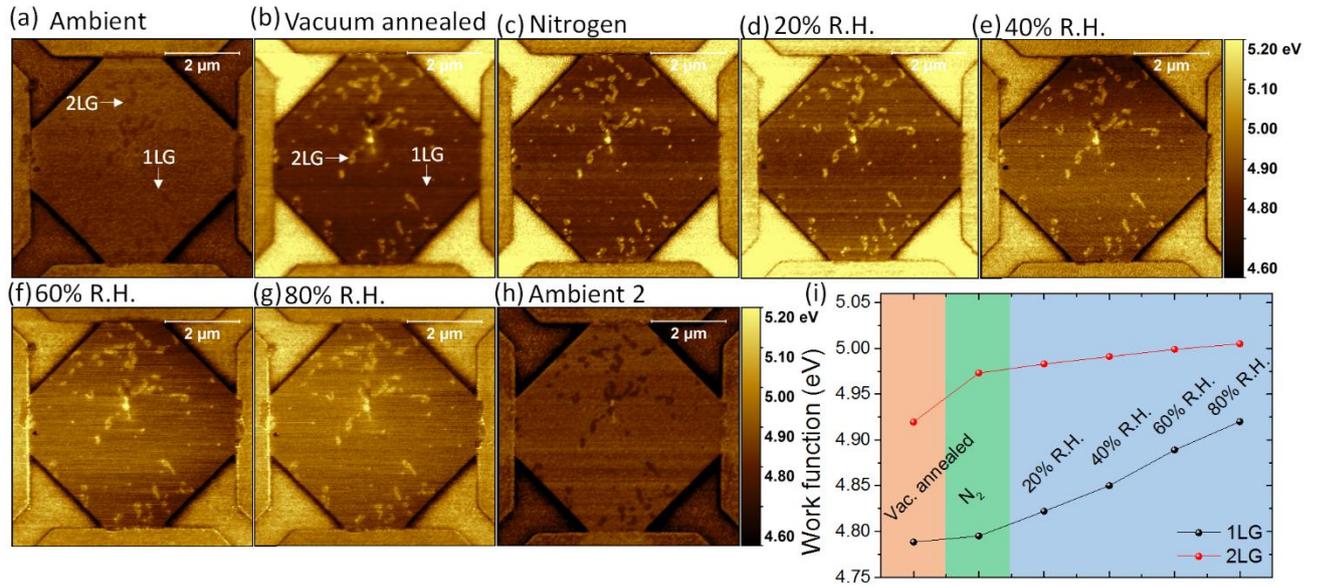

**Figure 4:** Work function maps of QFS 1LG on SiC in different environmental conditions: (a) ambient, (b) following annealing in vacuum, (c) nitrogen, (d-g) humidity levels of 20-80% R.H. and (h) re-exposure to ambient air. The active area of the graphene van der Pauw structure features 1LG with 2LG island inclusions. (i) Absolute values of work function measurements for the QFS 1LG (black) and 2LG (red).

Similarly to the as-grown sample, exposure to N$_2$ results in the increase of work functions of 1LG and 2LG to $\Phi_{1LG} = 4.8\ eV$ and $\Phi_{2LG} = 4.97\ eV$, respectively. At 20% R.H. the p-doping effect of water QFS 1LG becomes apparent, with the work function of both 1LG and 2LG increasing, $\Phi_{1LG} = 4.82\ eV$ and $\Phi_{2LG} = 4.98\ eV$, respectively, with the 1LG being affected to larger extend than the 2LG (see figure 4i). The difference in work function between the two layers at 20% R.H. was decreasing to $\Delta\Phi_{2-1LG} = 160\ meV$. This suggests that even in the case of the quasi-free standing graphene, the two layers exhibit different response to water because of the higher hydrophobicity of 2LG compared to 1LG, a result which is consistent with previous reports for *AB*-stacked graphene, stating that thicker graphene screens the electrostatic potential of the substrate, therefore decreased water-graphene interactions[7,8,38]. Further increase of humidity up to 80% R.H. results in a continuous increase of the work function for both layers ($\Phi_{1LG} = 4.92\ eV$ and $\Phi_{2LG} = 5.01\ eV$, respectively), with the difference reaching $\Delta\Phi_{2-1LG} = 90\ meV$. Contrary to the as-grown sample, where the work function contrast between the two layers was reversed at 80% R.H., for the QFS 1LG it still remains unchanged (*i.e.* $\Delta\Phi_{2-1LG} > 0$) in these environmental conditions. Succeeding re-exposure of the sample in ambient air, the work function difference between the two layers recovers to $\Delta\Phi_{2-1LG} = -90\ meV$. Similarly to the as-grown graphene, even at high humidity levels, $\Delta\Phi_{2-1LG}$ did not reach the initial values, indicating that additional airborne contaminants are also responsible for the high p-doping of the sample measured in ambient conditions[8].

## 4. Conclusion

In summary, the electronic properties of QFS 1LG were investigated in controlled humidity environments and compared to as-grown graphene on SiC. Annealing of both as-grown and QFS 1LG in vacuum, resulted in desorption of loosely bound p-dopants originating from ambient air and consequent decrease (increase) in hole (electron) concentration in QFS 1LG (as-grown). Furthermore, we demonstrated that water vapour acts as a p-dopant on both as-grown and QFS 1LG, with the latter exhibiting a two-fold greater change in carrier concentration, compared to the as-grown graphene on SiC(0001). We explain the enhanced sensitivity of QFS 1LG to water by the p-type nature of the sample, which allows water molecules to orient in a way that lowers the hydrophobicity in agreement with previous findings. Accordingly, higher effective doping is achieved. The direct consequence of the enhanced charge carrier and short-range impurity scattering, owed to the presence of a water layer, is ~5% mobility degradation in QFS 1LG (comparing the nitrogen stage to 80% R.H.). However, even in the extremely high humidity environment (80% R.H.), the superior carrier mobility of QFS 1LG compared to as-grown graphene is retained (*i.e.*, more than three times higher).

Furthermore, using KPFM we studied the local thickness-dependent effect of water on electronic properties of both types of graphene. We demonstrate a more pronounced change in work function of 1LG compared to 2LG under the same water exposure, a result which is in a good agreement with previous transport measurements on as-grown graphene. Nevertheless, the work function difference between 1LG and 2LG in both as-grown and QFS even at 80% R.H. does not reach the differences observed in ambient air, indicating that additional airborne contaminants are also responsible for the high p-doping of the sample measured in ambient conditions.

We successfully demonstrate for the first time the influence of water vapour on the electronic properties of QFS 1LG, by highlighting the role of substrate-induced doping in graphene on SiC(0001) and subsequently the adsorption of water. This concludes that the effects of water vapour on the electronic properties of both as-grown and QFS 1LG can be controlled by appropriate engineering of the substrate and the graphene-substrate interface. These results need to be taken into account in the development of stable graphene based electronics, such as GFETs, future calibrated sensors and metrology standards to be used in sensing.

## 5. Acknowledgments

Authors acknowledge support of EC grants Graphene Flagship (No. CNECT-ICT-604391), EMRP under project GraphOhm (No. 117359) and NMS under the IRD Graphene Project (No. 119948). The work was carried out as part of an Engineering Doctorate program in Micro- and NanoMaterials and Technologies, financially supported by the EPSRC under the grant EP/G037388, the University of Surrey and the National Physical Laboratory. The authors would like to acknowledge Alexander Tzalenchuk and Ivan Rungger for all the useful discussions.